\documentclass[aps,prb,twocolumn,superscriptaddress]{revtex4}

\usepackage{amsmath}
\usepackage{graphicx}
\usepackage{dcolumn}
\usepackage{bm}

\begin{document}

\title{Deviation from Fermi-liquid behavior in two-dimensional surface states of Au-induced nanowires on Ge(001) by correlation and localization}

\author{Koichiro Yaji}
\email{yaji@issp.u-tokyo.ac.jp}
\affiliation{Institute for Solid State Physics, The University of Tokyo, 5-1-5 Kashiwanoha, Kashiwa, Chiba 277-8581, Japan}

\author{Ryu Yukawa}
\affiliation{Institute for Solid State Physics, The University of Tokyo, 5-1-5 Kashiwanoha, Kashiwa, Chiba 277-8581, Japan}

\author{Sunghun Kim}
\affiliation{Institute for Solid State Physics, The University of Tokyo, 5-1-5 Kashiwanoha, Kashiwa, Chiba 277-8581, Japan}

\author{Yoshiyuki Ohtsubo}
\affiliation{Synchrotron SOLEIL, L'Orme des Merisiers, Saint-Aubin-BP 48, F-91192 Gif sur Yvette, France}

\author{Patrick Le F\`{e}vre}
\affiliation{Synchrotron SOLEIL, L'Orme des Merisiers, Saint-Aubin-BP 48, F-91192 Gif sur Yvette, France}

\author{Fran\c{c}ois Bertran}
\affiliation{Synchrotron SOLEIL, L'Orme des Merisiers, Saint-Aubin-BP 48, F-91192 Gif sur Yvette, France}

\author{Amina Taleb-Ibrahimi}
\affiliation{Synchrotron SOLEIL, L'Orme des Merisiers, Saint-Aubin-BP 48, F-91192 Gif sur Yvette, France}

\author{Iwao Matsuda}
\affiliation{Institute for Solid State Physics, The University of Tokyo, 5-1-5 Kashiwanoha, Kashiwa, Chiba 277-8581, Japan}

\author{Kan Nakatsuji}
\affiliation{Department of Materials Science and Engineering, Tokyo Institute of Technology, J1-3, 4259 Nagatsuta-cho, Midori-ku, Yokohama, Kanagawa 226-8502, Japan}

\author{Fumio Komori}
\email{komori@issp.u-tokyo.ac.jp}
\affiliation{Institute for Solid State Physics, The University of Tokyo, 5-1-5 Kashiwanoha, Kashiwa, Chiba 277-8581, Japan}

\begin{abstract}

The electronic states of Au-induced atomic nanowires on Ge(001) (Au/Ge(001) NWs) have been investigated by angle-resolved photoelectron spectroscopy with linearly polarized light. 
We have found three electron pockets around $\bar{J}$$\bar{K}$, where the Fermi surfaces are closed in a surface Brillouin zone, indicating that the surface states of Au/Ge(001) NWs are two-dimensional whereas the atomic structure is one-dimensional. 
The two-dimensional metallic states exhibit remarkable suppression of the photoelectron intensity near a Fermi energy. 
This suppression can be explained by the correlation and localization effects in disordered metals, which is a deviation from a Fermi-liquid model.  

\end{abstract}


\maketitle

Interacting electrons in two- and three-dimensional metals without disorder are well described by Fermi-liquid theory, where low-energy excitations of the interacting electrons can be treated as non-interacting quasi particles. 
The interaction between electrons in one-dimensional systems breaks down the Fermi-liquid theory and is described as a Tomonaga-Luttinger liquid (TLL) \cite{Luttinger_63, Voit_95, Voit_00}. 
Atomic nanowires (NWs) formed on semiconductor surfaces have been regarded as suitable systems to explore such intriguing physics of a TLL in real materials \cite{Yeom_99, Segovia_99, Losio_01, Crain_03, Nakatsuji_09, Blumenstein_11_1, Meyer_11, Nakatsuji_11_PRB, Yaji_13}. 
While the atomic structures of the NWs have been well established on the semiconductor surfaces, the TLL behavior has not yet been demonstrated experimentally \cite{Yeom_99, Losio_01, Crain_03, Yaji_13} except in gold-induced atomic NWs on Ge(001) (Au/Ge(001) NWs) \cite{Blumenstein_11_1, Meyer_11}. 
One of the difficulties is that the electrons in these NWs are not always strictly one-dimensional; two-dimensional undulation of the Fermi surface, reflecting inter-chain interaction, was often observed as in In/Si(111)-(4$\times$1) \cite{Yeom_99} and Au/Si(553) \cite{Crain_03} by angle-resolved photoelectron spectroscopy (ARPES). 
Another difficulty is that lattice disorder in the NWs at the surfaces makes the TLL behavior smeared out \cite{Starowicz_02}. 
Moreover, correlation and localization effects in disordered metals induce non-Fermi-liquid behavior \cite{Lee_85}, such as the decrease of density of state (DOS) around Fermi energy ($E_{\rm F}$) \cite{Altshuler_79, McMillan_81_PRB}. 

For Au/Ge(001) NWs, a TLL state was claimed on the basis of the decreasing DOS with a power law of the binding energy near the Fermi energy ($E_{\rm F}$), as observed both in the tunneling spectra by scanning tunneling microscope (STM) and the integrated ARPES intensity \cite{Blumenstein_11_1, Meyer_11}. 
However, it has been reported that the metallic surface band of Au/Ge(001) NWs exhibits two-dimensional nature \cite{Nakatsuji_09, Nakatsuji_11_PRB}, and the direction of the surface Brillouin zone (SBZ) is still under discussion \cite{Nakatsuji_11_NP}. 
In addition, a unique lattice disorder at the surface in a long range has been observed by STM. \cite{Niikura_11, Blumenstein_11_2, Safaei_13, Blumenstein_13, SM_disorder} 

In this Rapid Communication, we report on the electronic states of Au/Ge(001) NWs studied by high-resolution ARPES using linearly polarized light. 
We show that the Fermi surfaces of Au/Ge(001) NWs are closed in SBZ and the surface states are undoubtedly two-dimensional, indicating that the interpretation as the TLL state is failed. 
Nevertheless, we found that the two-dimensionally-integrated ARPES intensity of each surface band shows suppression of DOS with decreasing the binding energy to $E_{\rm F}$. 
We attribute this to the correlation and localization effects in disordered metals \cite{Lee_85, Altshuler_79, McMillan_81_PRB}. 
The lattice disorder in the long range inherent to Au/Ge(001) NWs induces the localization of the interacting surface electrons with keeping the band electron character. 

\begin{figure}
\includegraphics{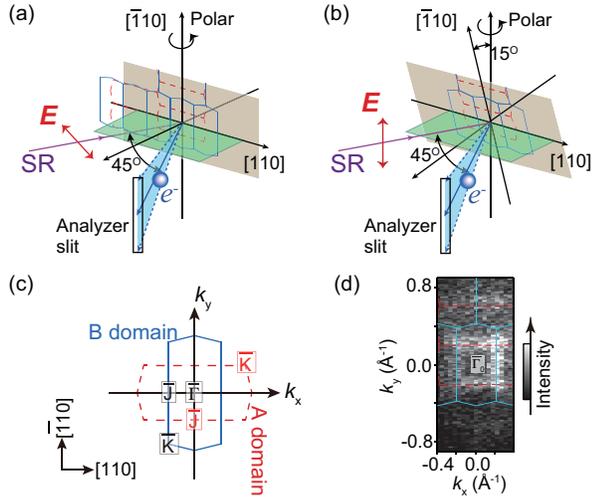}
\caption{\label{fig:epsart}
(Color online) 
(a, b) Experimental geometries for ARPES, labeled the LH geometry (a) and the LV geometry (b), where the angle between SR and the analyzer is fixed to 45$^\circ$. 
Light green parallelograms represent the SR incident planes. 
(c) Definition of the SBZ's of the double-domain sample of Au/Ge(001) NWs in the present study. 
For A (B) domain, the nanowires are aligned parallel to the [110] ([$\bar{1}$10]) direction. 
(d) Constant-energy ARPES intensity map at $E_{\rm B}$ = 20 meV recorded with $h\nu$ = 96 eV with the LH geometry. 
}
\end{figure}

Our ARPES experiments were performed at CASSIOP\'{E}E beamline in synchrotron SOLEIL. 
The sample temperature was 6 K during the measurements. 
The experimental geometries are represented in Figs. 1(a) and 1(b). 
We used synchrotron radiation (SR) with linear horizontal (LH) and linear vertical (LV) polarization, where the electric-field vectors of LH and LV are parallel and perpendicular to a light incident plane, respectively. 
A slit of a hemispherical electron energy analyzer (VG Scienta R4000) is perpendicular to the light incident plane. 
For the LH geometry in Fig. 1(a), the electric-field vector of the incident light coincides with a ($\bar{1}$10) plane of a sample. 
For the LV geometry in Fig. 1(b), the sample was tilted by an angle of 15$^\circ$ around [110] in order to acquire the photoelectrons from 4th SBZ of Au/Ge(001)-{\it c}(2$\times$8). 
Intensity mapping of ARPES was performed with rotating the sample around the polar axis shown in the figure. 
The energy resolution for the measurement with the photon energy of 31 eV was set to 20 meV. 

The Au/Ge(001) NWs were prepared {\it in situ} in the molecular beam epitaxy chamber connected to the ARPES chamber. 
We used a flat {\it n}-type Ge(001) substrate, which provides double-domain samples of Au/Ge(001) NWs with the {\it c}(2$\times$8) reconstruction. 
The clean surface was prepared with several cycles of 0.8 keV Ar$^{+}$ bombardment at 670 K and subsequent annealing up to 900 K for a few minutes. 
Au was deposited onto the clean surface kept at 740 K from a tungsten filament coated by Au, which gave a sharp and low-background {\it c}(2$\times$8) low-energy electron diffraction  pattern \cite{SM_disorder}. 
In Fig. 1(c), the SBZs of the double domain sample, labeled A and B domains, are represented. 
Here, we define the $k_{\rm x}$ and $k_{\rm y}$ to be parallel to the [110] and [$\bar{1}$10] directions, respectively. 
A constant-energy ARPES map of the reconstructed surface with $h\nu$ = 96 eV is shown in Fig. 1(d), and is quite similar to the previous report \cite{Meyer_11}. 

\begin{figure}
\includegraphics{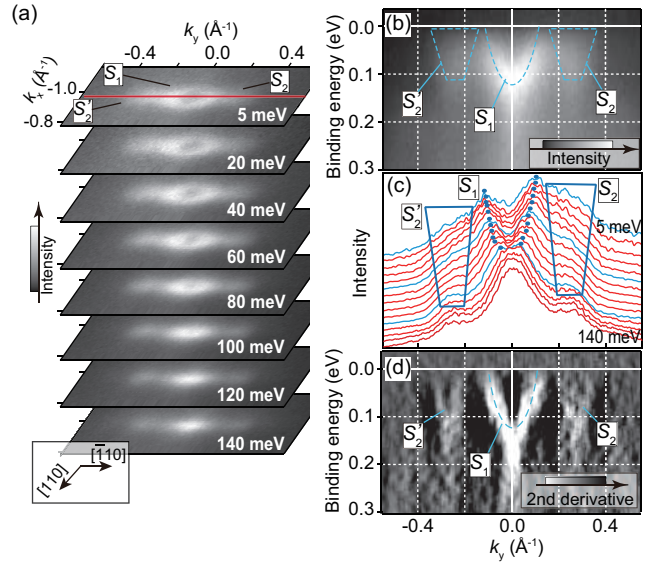}
\caption{\label{fig:epsart}
(Color online) 
(a) Constant-energy ARPES intensity maps at 5--140 meV recorded with $h\nu$ = 31 eV with the LH geometry, where each image is obtained by integration of the photoelectron intensity within a 5 meV energy window centered at each binding energy. 
(b) ARPES intensity image taken along the red line in (a). 
(c) MDCs obtained from (b) at 5--140 meV with integration of the photoelectron intensity within a 10 meV energy window, where each MDC is normalized to its maximum. 
(d) Second-derivative image calculated from (b), where the second derivative is calculated for MDCs. 
}
\end{figure}

Figure 2(a) shows ARPES constant energy maps at the binding energies between 5 meV and 140 meV recorded with the photon energy of 31 eV and the LH geometry. 
Here, the photoelectron intensities from the surface-state bands of Au/Ge(001) NWs are enhanced around $k_{\rm x}$ = $-$0.98 \AA$^{-1}$ with $h\nu$ = 31 eV \cite{SM_ARPES_wide_range}. 
The constant energy map at 5 meV, just below $E_{\rm F}$, can be regarded as a Fermi surface map because the energy is rather small compared with the energy resolution. 
Figures 2(b) and 2(d) show the ARPES intensity map and its second derivative recorded along a line shown in Fig. 2(a) with the LH geometry. 
Momentum distribution curves (MDCs) obtained from Fig. 2(b) are represented in Fig. 2(c).  

The Fermi surface of $S_{\rm 1}$ is closed in the SBZ. 
The $S_{\rm 1}$ band has upward energy dispersion from $k_{\rm y}$ = 0 \AA$^{-1}$ and clearly crosses the Fermi level on $k_{\rm y}$. 
Such energy dispersion of $S_{\rm 1}$ was also observed with the photon energy of 96 eV \cite{SM_96eV}. 
We emphasize that the surface-state bands should not cross the Fermi level along $k_{\rm y}$ if the surface states show one-dimensional nature and trough-like energy dispersion in the SBZ as reported in the references of 10 and 21. 
The $S_{\rm 1}$ band also crosses the Fermi level on $k_{\rm x}$ (see Fig. 3(b)). 
Thus the $S_{\rm 1}$ band is two-dimensional at $E_{\rm F}$.

In the area of $S_{\rm 2}$ ($S'_{\rm 2}$), we can recognize finite ARPES intensity, even at the binding energy of 5 meV. 
In the mapping images and MDCs, the diameter of the round ARPES intensity at the $S_{\rm 2}$ ($S'_{\rm 2}$) area increases with decreasing the binding energy toward $E_{\rm F}$. 
This suggests that surface bands with upward energy dispersion cross the Fermi level on the $k_{\rm y}$ axis in these areas. 
The possible $S_{\rm 2}$ and $S'_{\rm 2}$ bands also cross the Fermi level on the direction parallel to $k_{\rm x}$ (see Fig. 3(c)).
These results clearly contradict with the interpretation of a one-dimensional band \cite{Blumenstein_11_1, Meyer_11}. 
The widths of the MDCs ought to be the same if the band is one-dimensional.

Consequently, the surface-state bands $S_{\rm 1}$, $S_{\rm 2}$ and ($S'_{\rm 2}$) are all two-dimensional\cite{Nakatsuji_05}, and the Fermi surfaces of Au/Ge(001) NWs consists of three oval-shaped electron pockets elongated in the $k_{\rm y}$ direction. 
A center of the Fermi surface for $S_{\rm 1}$ is located at ($k_{\rm x}$, $k_{\rm y}$) = (-0.98, 0) \AA$^{-1}$, where the wave vector of 0.98 \AA$^{-1}$ corresponds to five times of the distance of $\bar{\Gamma}$$\bar{J}$. 
Centers of the Fermi surfaces of $S_{\rm 2}$ and $S'_{\rm 2}$ are symmetrically shifted from $k_{\rm y}$ = 0 \AA$^{-1}$ by $\pm$0.26 \AA$^{-1}$ on $k_{\rm x}$ = 0.98 \AA$^{-1}$. 

\begin{figure*}
\includegraphics[width=160mm]{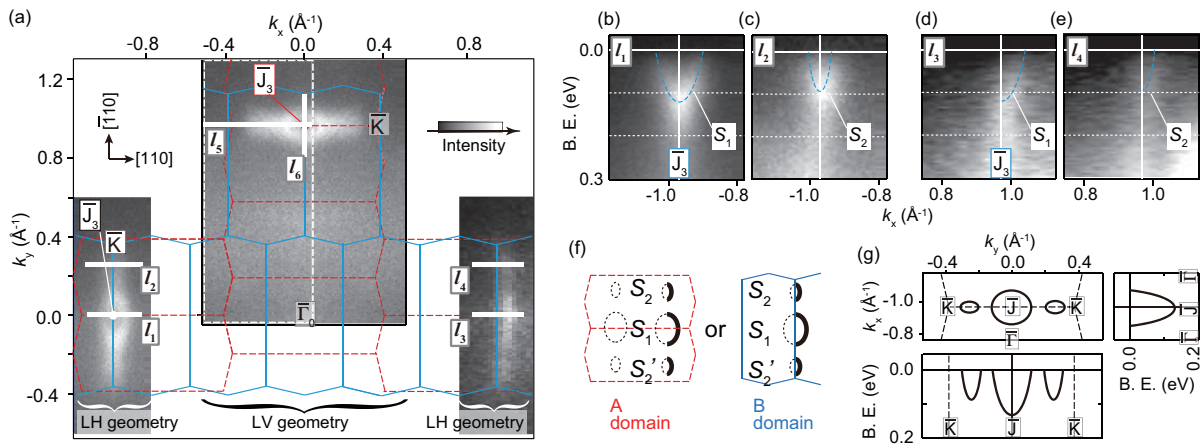}
\caption{\label{fig:epsart}
(Color online) 
(a) Constant-energy ARPES intensity maps obtained by integration of the photoelectron intensities within a 5 meV energy window centered at 5 meV. 
For the map with the LV geometry, a thin dashed rectangle represents a region actually measured, and then the rest of the image was obtained by mirror symmetry operations.
(b-e) ARPES intensity maps taken along bold solid lines, $l_{\rm 1}$--$l_{\rm 4}$, shown in (a). 
(f) Fermi surfaces schematically drawn together with the SBZs of the A and B domains. 
Bold solid and half ovals represent the schematics of the Fermi surface experimentally observed. 
Thin dashed ovals represent the Fermi surface of which the photoelectron intensities are diminished. 
The tentative Fermi surfaces based on the A domain are represented in the left panel while the Fermi surfaces belong to the B domain in the right panel. 
(g) Schematic drawing of the Fermi surfaces and the band structures of $S_{\rm 1}$, $S_{\rm 2}$ and $S'_{\rm 2}$. 
}
\end{figure*}

The above is mentioned without stipulating the SBZ. 
Hereafter, we start to discuss the relationship between the surface-state bands and the SBZ. 
The ARPES intensities of the metallic bands $S_{\rm 1}$ and $S_{\rm 2}$ from the present double-domain sample largely depend on the light polarization of 31 eV. 
Figure 3(a) shows constant-energy intensity maps at the binding energy of 5 meV. 
Here, the maps in the range of 0.8 $<$ $k_{\rm x}$ $<$ 1.2 \AA$^{-1}$ and $-$1.2 $<$ $k_{\rm x}$ $<$ $-$0.8 \AA$^{-1}$ were measured with the LH geometry, and that in the range of $-$0.5 $<$ $k_{\rm x}$ $<$ +0.5 \AA$^{-1}$ with the LV geometry. 
For the maps with the LH geometry, the photoelectron intensities are prominent only around $k_{\rm x}$ $\sim$ $\pm$0.98 \AA$^{-1}$ and $-$0.35 $<$ $k_{\rm y}$ $<$ +0.35 \AA$^{-1}$ in the area of the $k$-space shown in Fig. 3(a). 
Figures 3(b)--3(e) exhibit the band structures recorded along $l_{\rm 1}$--$l_{\rm 4}$ shown in Fig. 3(a). 
We observed upward energy dispersions and the oval-shaped Fermi surfaces for $S_{\rm 1}$, $S_{\rm 2}$ and $S'_{\rm 2}$ on a negative $k_{\rm x}$.  
The photoelectron intensity from each band is almost symmetric with respect to $k_{\rm x}$ = $-$0.98 \AA$^{-1}$.
On the other hand, for the positive $k_{\rm x}$, the photoelectron intensities from $S_{\rm 1}$ and $S_{\rm 2}$ are visible only at $k_{\rm x}$ larger than +0.98 \AA$^{-1}$. 
Concerning the Fermi surfaces, half ovals are consistently observed at $k_{\rm x}$ $\geq$ +0.98 \AA$^{-1}$. 

We schematically draw possible Fermi surfaces of these bands on the positive $k_{\rm x}$ together with the SBZs of the A and B domains in Fig. 3(f). 
If the surface bands belong to the B domain, the band centers are on the boundary of SBZs. 
Then, the intensities from $S_{\rm 1}$, $S_{\rm 2}$ and $S'_{\rm 2}$ are enhanced in the 4th SBZ, and are suppressed in the 3rd SBZ. 
This modulation of the photoelectron intensities can be attributed to the interference effect in the photoemission process, known as Brillouin-zone-selection effect  \cite{Shirley_95, Daimon_95, Gierz_11}.
We note that the incident light polarization is an important factor in the photoemission process and modifies the interference effect. 
In the present case, the intensity modulation is conspicuous only at $k_{\rm x}$ $\sim$ $+$0.98. 
This can be explained by the difference of the polarization \cite{comment_1}. 
If we assume that observed bands belong to the A domain, the photoelectron intensities from them are suddenly diminished in the same SBZ at $k_{\rm x}$ $\leq$ +0.98 \AA$^{-1}$, as shown in a left panel of Fig. 3(f). 
As far as we know, there is no theory to account such change of the photoemission matrix element and thus the angular distribution of photoelectron intensity. 
Here, we conclude that the surface metallic bands are around $\bar{J}$$\bar{K}$ and each electron pocket exhibits symmetric energy dispersion with respect to $\bar{J}\bar{K}$. 
The fixed band structures of Au/Ge(001) NWs are schematically depicted in Fig. 3(g). 
This conclusion is consistent with the previous low-resolution ARPES result using a single-domain sample \cite{Nakatsuji_11_PRB}. 
The present high-resolution ARPES with the optimized photon energy of $h\nu$ = 31 eV provides clear information on the band structure of Au/Ge(001) NWs compared with the previous one with a He discharge lamp ($h\nu$ = 21.2 eV). 
We note that the assignment of the SBZ in the references of 10 and 21 was inappropriate. 
For correcting it, the SBZ should have been rotated by 90$^{\circ}$. 

From above discussion, $S_{\rm 1}$ and $S_{\rm 2}$ around $k_{\rm x}$ $\sim$ $\pm$0.98 \AA$^{-1}$ and $-$0.35 $<$ $k_{\rm y}$ $<$ +0.35 \AA$^{-1}$ observed with the LH geometry belong to the B domain. 
We note that $S_{\rm 1}$ and $S_{\rm 2}$ in this area could not be detected with the LV geometry. 
On the other hand, $S_{\rm 1}$ and $S_{\rm 2}$ were found with the LV geometry only around $-$0.35 $<$ $k_{\rm x}$ $<$ $+$0.35 \AA$^{-1}$ and $k_{\rm y}$ $\sim$ $+$0.98 \AA$^{-1}$, which corresponds to the $\bar{J}\bar{K}$ axis of the A domain, as shown in Fig. 3(a). 
These surface bands thus belong to the A domain. 
We observed the upward energy dispersions for $S_{\rm 1}$ and $S_{\rm 2}$ of the A domain along $l_{\rm 5}$ shown in Fig. 3(a) \cite{SM_ARPES_LV}, being consistent with the result shown in Figs. 2(b) and 2(c). 

According to dipole selection rule, initial state with even (odd) symmetry with respect to the ($\bar{1}$10) plane is observable with the LH (LV) light polarization. 
We thus assign that the surface bands of the B domain predominantly exhibit the even symmetry on the ($\bar{1}$10) plane. 
This result should be considered to analyze the symmetry of the surface states by band-structure calculations. 

\begin{figure}
\includegraphics{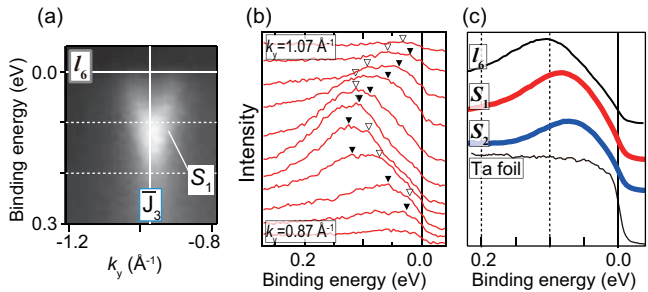}
\caption{\label{fig:epsart}
(Color online) 
(a) ARPES intensity map taken with the LV geometry along $l_{\rm 6}$ shown in Fig. 3(a).
(b) EDCs of the ARPES image shown in (a). 
Triangle symbols represent peak positions of $S_{\rm 1}$. 
(c) The top one represents the summation of EDCs shown in (b). 
The second and third from the top are two-dimensionally integrated ARPES intensities of $S_{\rm 1}$ and $S_{\rm 2}$ over the area of each band in the SBZ, respectively.  
The bottom one is the Fermi edge taken from Ta foil electrically touched to the sample. 
}
\end{figure}

Figure 4(b) shows energy distribution curves (EDCs) obtained from the ARPES intensity map shown in Fig. 4(a). 
We can recognize that $S_{\rm 1}$ is split into two, suggesting that $S_{\rm 1}$ exhibits the Rashba-type band splitting \cite{Rashba_60, Bychkov_84, SM_Rashba}. 
The summation of these EDCs is shown in Fig. 4(c). 
The line-shape of the EDCs is in good agreement with the result reported by Meyer {\it et al.} \cite{Meyer_11}. 
Note that this summation does not provide the DOS of the two-dimensional band.  

The two-dimensionally integrated constant-energy ARPES intensity of each band, $S_{\rm 1}$ and $S_{\rm 2}$, over the area of each band in the SBZ, corresponds to the DOS if we neglect the matrix element effect in the photoemission process. 
The results are shown in Fig. 4(c). 
The integrated photoelectron intensities from both $S_{\rm 1}$ and $S_{\rm 2}$ are much suppressed near $E_{\rm F}$ compared with a spectrum taken from a Ta foil, regarded as the Fermi-Dirac distribution of a constant DOS. 
The result is quite in contrast to no suppression of the photoelectron intensity at $E_{\rm F}$ in Pt/Ge(001) NWs with little lattice disorder \cite{Yaji_13}.

The suppression of the photoelectron intensity at $E_{\rm F}$ was claimed as evidence of the TLL behavior previously \cite{Blumenstein_11_1, Meyer_11}. 
However, the surface states are two-dimensional, as mentioned above. 
For discussion on the origin of the observed low intensity of photoelectrons near $E_{\rm F}$, we should consider in the matrix element effect in the photoemission process in principle. 
In the present system, however, the effect must be small and the observation surely indicates the decrease of DOS with decreasing the binding energy toward $E_{\rm F}$ because the suppression of DOS was also observed by the scanning tunneling spectroscopy (STS) \cite{Blumenstein_11_1}. 

The suppression of DOS around $E_{\rm F}$ can be attributed to the correlation and localization effects in disordered metals \cite{Altshuler_79, McMillan_81_PRB, Lee_85}. 
A small decrease of DOS at $E_{\rm F}$ due to this effect in a so-called weakly-localized metal continuously develops with increasing the disorder, and finally an energy gap is formed at the point of the metal-insulator transition. 
This anomaly of DOS near $E_{\rm F}$ in the disordered metallic system is interpreted as a precursor effect of the opening of the correlation gap in the insulator phase \cite{McMillan_81_PRB}. 
The suppression of DOS around $E_{\rm F}$ with the increase of disorder was actually reported in the tunneling spectra of amorphous Au$_{1-x}$Ge$_{x}$ \cite{McMillan_81_PRL}, Nb$_{1-x}$Si$_{x}$\cite{Hertel_83} and integrated photoelectron intensity from the valence states of granular Pd \cite{Weng_85}. 

In the present two-dimensional system, surface-state electrons are considerably scattered owing to the unique disorder in the surface lattice, which was commonly observed in the atomic images by STM \cite{Niikura_11, Blumenstein_11_2, Safaei_13, Blumenstein_13, SM_disorder}. 
The non-periodic potential for the two-dimensional electrons induces the DOS suppression while the weak disorder can keep the surface band from serious broadening. 
In the theory \cite{ McMillan_81_PRB, Lee_85}, the DOS suppression recovers with increasing temperature. 
This was actually observed in the previous STS experiments of Au/Ge(001) NWs \cite{Blumenstein_11_1}. 

At present, however, the formula expressing the DOS as a function of energy, which depends on the nature of the electron-electron interaction, on the universality class related to the strength of spin-orbit interaction, and on the degree of disorder \cite{Lee_85}, is not clear for the present two-dimensionally interacting system \cite{SM_DOS_fit}. 
Further theoretical studies as well as systematic experimental studies such as on the same system with different disorder are necessary for the quantitative analysis of the correlation and localization effects. 

In summary, we have investigated the electronic band structure of Au/Ge(001) NWs by ARPES with linearly polarized light. 
We observed three electron pockets on $\bar{J}\bar{K}$, where the Fermi surfaces are closed in the SBZ. 
This is direct evidence of the two-dimensional metallic states in this system. 
The polarization dependence of ARPES provided conclusive information to fix the direction of the SBZ in the double-domain samples. 
For the two-dimensional surface states, we observed remarkable suppression of the photoelectron intensity just below $E_{\rm F}$. 
This suppression can be driven by the correlation and localization effects due to the unique lattice disorder at the surface, which is a key mechanism to understand the deviation from the Fermi-liquid model in real materials. 

The authors thank Daniel Ragonnet and Fran\c{c}oise Deschamps for their support during the experiments at the CASSIOP\'{E}E beamline in synchrotron SOLEIL. 
They also thank Han Woong Yeom for fruitful discussion on the experimental results.
The present work was financially supported by the JSPS Grant-in-Aid for Young Scientists (B), Grant No. 24740197 and for Scientific Research (B), Grant No. 26287061.


\begin{references}
\bibitem{comment_1} The incident angles of SR to acquire the photoelectrons from the wave vectors of $k_{\rm x}$ $\sim$ $-$0.98 \AA$^{-1}$ and $k_{\rm x}$ $\sim$ +0.98 \AA$^{-1}$ are $\sim$23$^\circ$ and $\sim$67$^\circ$ with respect to the surface normal, respectively. Note that the angle between SR and the analyzer is fixed to 45$^\circ$.  Thus, the polarization conditions are markedly different between $k_{\rm x}$ $\sim$ $\pm$0.98 \AA$^{-1}$, where the vertical component of the electric-field vector of the LH polarization on the surface is about 40$\%$ for $k_{\rm x}$ $\sim$ $-$0.98 \AA$^{-1}$, while the vertical component of the electric-field vector more than 90$\%$ for $k_{\rm x}$ $\sim$ +0.98 \AA$^{-1}$.
\end{references}
\end{document}